\documentclass[11pt]{article}
\usepackage{amsmath}
\usepackage{amssymb}
\usepackage{graphicx}

\def\be{\begin{equation}}
\def\ee{\end{equation}}
\def\bea{\begin{eqnarray}}
\def\eea{\end{eqnarray}}

\begin{document}

\begin{titlepage}
\begin{flushright}
ECT$^*$-07-14\\
HD-THEP-07-15\\
\end{flushright}
\vfill
\begin{center}
\boldmath
{\LARGE{\bf Diffractive Neutral Pion Production,}}\\[.2cm]
{\LARGE{\bf Chiral Symmetry and the Odderon\,$^*$}}
\unboldmath
\end{center}
\vspace{1.2cm}
\begin{center}
{\bf \Large
Carlo Ewerz\,$^{a,1}$, Otto Nachtmann\,$^{b,2}$
}
\end{center}
\vspace{.2cm}
\begin{center}
$^a$
{\sl
ECT\,$^*$, Strada delle Tabarelle 286, 
I-38050 Villazzano (Trento), Italy}
\\[.5cm]
$^b$
{\sl
Institut f\"ur Theoretische Physik, Universit\"at Heidelberg\\
Philosophenweg 16, D-69120 Heidelberg, Germany}
\end{center}                                                                
\vfill
\begin{abstract}
\noindent
We discuss the diffractive photo- and electroproduction of a single 
neutral pion at high energies where it can occur due to  odderon exchange. 
We show that this process is dynamically suppressed as a consequence of  
chiral symmetry. Our result  reconciles earlier theoretical expectations with 
the non-observation of this reaction at HERA. 
\vfill
\end{abstract}
\vspace{5em}
\hrule width 5.cm
\vspace*{.5em}
{\small \noindent
$^1$ email: Ewerz@ect.it \\
$^2$ email: O.Nachtmann@thphys.uni-heidelberg.de\\
$^*$ Based on invited talks given by C.E.\ at 
XVth International Workshop on Deep-Inelastic Scattering and 
Related Subjects (DIS 2007), M\"unchen, April 2007, 
and by O.N.\ at 
12th International Conference on Elastic and Diffractive Scattering 
(EDS07), DESY Hamburg, May 2007.
}

\end{titlepage}

\section{Diffractive neutral pion production as a probe of the odderon}

In this talk we want to present the main results of 
our study \cite{Ewerz:2006gd} of diffractive production of 
a single neutral pion in photon-proton scattering at high energy, 
\begin{equation}
\label{reaction}
\gamma^{(*)}(q)+p(p) \, \longrightarrow \, \pi^0(q')+X(p')\,,
\end{equation}
where the photon can be real or virtual, and $X$ can be any 
diffractively produced hadronic system. For simplicity we will 
in the following assume that $X$ is a proton, but our considerations 
can also be applied to other states $X$, for example for $X=N^*$ 
or $X= n + \pi^+$ \cite{Ewerz:2006gd}. 
\begin{figure}[ht]
\centerline{\includegraphics[width=7.5cm]{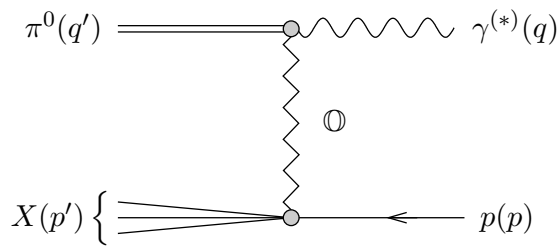}}
\caption{Diffractive photo- or electroproduction of a $\pi^0$ 
due to odderon exchange.}\label{Fig:diag}
\end{figure}
Since the photon and the neutral pion have opposite $C$-parity 
the object exchanged in this reaction must be odd under charge 
conjugation, and hence at high energy must be an odderon ($\mathbb O$), 
see Figure \ref{Fig:diag}. (Note that we draw the incoming particles 
to the right.) 

The odderon, the $C=-1$ partner of the pomeron, was introduced in 
\cite{Lukaszuk:1973nt,Joynson:1975az}, 
for a general review see \cite{Ewerz:2003xi}. 
It has since been studied in great detail especially 
from a theoretical point of view. But experimentally the odderon 
remains an elusive object. Some weak evidence for its existence has only 
been seen in elastic scattering at the ISR where the $pp$ and $p\bar{p}$ 
differential cross sections show a difference at around 
$|t|\sim 1.3\,\mbox{GeV}^2$ \cite{Breakstone:1985pe}, for a 
recent discussion see  \cite{Dosch:2002ai}. There, however, the odderon 
is only one among many contributions and hence difficult to pin down. 
In recent years it has been realized that the chances to observe the 
odderon are better in exclusive processes in which the odderon 
essentially gives the only contribution. 
As an important example of this strategy the reaction (\ref{reaction}) 
has been proposed and discussed in 
\cite{Schafer:1992pq,Barakhovsky:1991ra,Kilian:1997ew}. 

A detailed analysis based on a nonperturbative model of QCD dynamics 
performed in \cite{Berger:1999ca} led to the prediction 
\begin{equation}
\label{predict}
\sigma(\gamma p \to \pi^0 N^*) \approx 300 \,{\rm nb} \,,
\end{equation}
while the subsequent experimental search at HERA \cite{Adloff:2002dw} 
did not  find a signal and resulted in the upper bound 
\begin{equation}
\label{exbound}
\sigma(\gamma p \to \pi^0 N^*)  <  49\,{\rm nb} \,.
\end{equation}
Possible causes for the failure of the prediction of \cite{Berger:1999ca} 
were discussed in \cite{Donnachie:2005bu}. 
Since the reaction (\ref{reaction}) has the largest phase space 
of all processes in which hadrons are diffractively produced 
a strong dynamical suppression appeared necessary in order to 
provide a likely reason. In \cite{Ewerz:2006gd} we have found that 
the approximate chiral symmetry of QCD indeed induces a strong 
suppression. 

\section{PCAC}

Let us define a quark field operator describing up and down type 
quarks, $\psi(x) = (u(x), d(x))^{\rm T}$, and the associated triplet of 
axial vector currents ($a=1,2,3$)
\begin{equation}
A_\mu^a(x) = \bar{\psi}(x) \gamma_\mu \gamma_5 
\frac{\tau^a}{2} \psi(x). 
\end{equation}
The well-known PCAC relation states that the divergence 
of this axial vector current is related to a correctly normalised 
pion field operator $\phi^a$ by 
\begin{equation}
\label{PCAC}
\partial_\lambda A^{a \lambda} (x) = 
\frac{f_\pi m_\pi^2}{\sqrt{2}} \, \phi^a(x) \,,
\end{equation}
where $f_\pi \approx 130 \, \mbox{MeV}$ is the pion decay constant, 
see for example \cite{AdlerDashen}. 
Let us now consider along with diffractive pion production the 
corresponding production of an axial vector current $A^3$, 
\begin{eqnarray}
\label{reactionpi0}
\gamma^{(*)}(q,\nu)+p(p,s)\, &\longrightarrow& \,\pi^0(q')+p(p',s') \,,
\\
\label{reactionA3}
\gamma^{(*)}(q,\nu)+p(p,s) \, &\longrightarrow& \, A^3(q',\mu)+p(p',s')\,,
\end{eqnarray}
and let us denote the corresponding amplitudes by 
${\cal M}^\nu(\pi^0;q',p,q)$ and  ${\cal M}^{\mu\nu} (A^3;q',p,q)$, 
respectively, which we consider for 
$q^2 \le 0$ and $q'^2 \le m_\pi^2$. 
Using the PCAC relation (\ref{PCAC}) we can then 
express the former amplitude in terms of the latter via 
\begin{equation}
\label{pcacampli}
{\cal M}^\nu(\pi^0;q',p,q)=
\frac{2\pi m_p\sqrt{2}}{f_\pi m^2_\pi}\,
(-q'^2+m^2_\pi) \,iq'_\mu {\cal M}^{\mu\nu}(A^3;q',p,q) \,. 
\end{equation}

\section{Axial vector current production}

\begin{figure}[ht]
\centerline{\includegraphics[width=15cm]{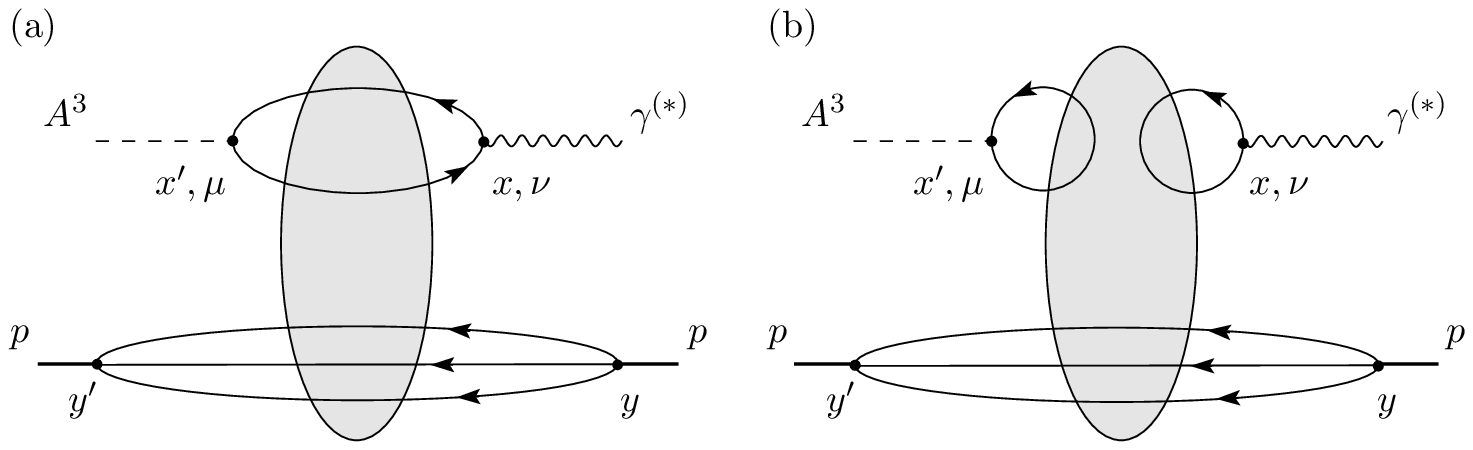}}
\caption{Leading diagrams at high energies for the 
reaction (\ref{reactionA3}).}
\label{Fig:fig3}
\end{figure}

The amplitude ${\cal M}^{\mu\nu}(A^3;q',p,q)$ for axial 
vector current production (\ref{reactionA3}) can be 
treated with the same general nonperturbative methods that 
were developed for Compton scattering in 
\cite{Ewerz:2004vf,Ewerz:2006vd}. 
We use the LSZ formula to relate the amplitude 
to Green's functions, and the latter can then be written as 
functional integrals over quark and gluon fields. Integrating 
out the quark degrees of freedom leads us to diagram classes 
characterised by their quark line skeleton. Those diagrams 
which contain the leading terms at high energies 
are shown in Figure \ref{Fig:fig3}, and they are exactly 
the odderon exchange diagrams on which we want to 
concentrate here. 
The solid lines represent full quark propagators in 
a given gluon field configuration, and the shaded blobs indicate 
the functional integral over the gluon fields. 

\section{From axial vector current to pion production}

We now consider the divergence of the amplitudes of Figure 
\ref{Fig:fig3}, that is we contract them with $q'^\mu$ or 
take a derivative $\partial^\mu$. For the axial vector current 
this gives at the quark level 
\begin{equation}
\label{delA3}
\partial^\mu A^3_\mu (x) = i [ m_u \bar{u}(x) \gamma_5 u(x) 
- m_d \bar{d}(x) \gamma_5 d(x) ] \,,
\end{equation}
and hence there is a factor $m_q$ of the light quark masses 
in the divergence amplitude. 
Note that the gluon anomaly does not contribute here. 
Such anomalous pieces are contained in the individual contribution 
of the quark flavours to the divergence of $A^3_\mu$ and would 
have the quark line topology of diagram (b) in Figure \ref{Fig:fig3}, 
but they cancel in $\partial^\mu A^3_\mu$. 

The quark loops in the diagrams of Figure \ref{Fig:fig3} 
which couple to the axial vector current contain 
a factor $\gamma_5$. As a consequence, these loops give rise 
to an additional factor $m_q$ of the light quark mass. This can 
be shown in a more detailed analysis which makes use of 
Lippmann-Schwinger equations for the quark propagator, 
for details see \cite{Ewerz:2006gd}. 
Hence we find that the divergence amplitude 
$q'_\mu{\cal M}^{\mu\nu}(A^3;q',p,q)$ is proportional to 
the square of the light quark masses. More precisely, 
\begin{equation}
\label{divamplim2}
q'_\mu{\cal M}^{\mu\nu}(A^3;q',p,q)
= m_u^2 \, {\cal C}^{(u)\nu}(q',p,q)
- m_d^2 \,{\cal C}^{(d)\nu}(q',p,q)
\,,
\end{equation}
where the functions $ {\cal C}^{(q)\nu}$ have pion poles but 
are otherwise finite. (These poles are cancelled by the explicit factor  
$(-q'^2+m^2_\pi)$ in (\ref{pcacampli}) when we insert 
(\ref{divamplim2}) there.) We know from the theory of chiral 
symmetry that the squared pion mass is linear in the light quark 
masses, 
\begin{equation}
m_\pi^2 = B (m_u + m_d)
\end{equation}
with 
\begin{equation}
B=-\frac{2}{f^2_\pi}\langle 0|\bar{u}(x)u(x)|0\rangle 
\approx 1800 \, \mbox{MeV}
\,.
\end{equation}
Therefore we can conclude from  (\ref{pcacampli}) and (\ref{divamplim2}) 
that the odderon exchange amplitude for pion production is 
proportional to the square of the pion mass, 
\begin{equation}
{\cal M}^\nu(\pi^0;q',p,q) 
\,\sim\, \frac{1}{m_\pi^2} \, {\cal M}^{\mu\nu} (A^3;q',p,q)
\,\sim \, \frac{1}{m_\pi^2} \, m_q^2 
\,\sim\, m_\pi^2 \,.
\end{equation}
We thus find that the odderon exchange amplitude for 
$\pi^0$-production vanishes in the chiral limit $m_\pi^2 \to 0$. 
This result can be generalised to the reaction (\ref{reaction}) 
with an arbitrary hadronic final state $X$. 

\section{Conclusion}

We have considered the diffractive process $\gamma^{(*)} p \to \pi^0 X$ 
at high energies where it should be dominated by odderon exchange. 
As a consequence 
of chiral symmetry the odderon exchange amplitude for this process 
vanishes in the chiral limit $m_\pi^2 \to 0$. We still expect a strong 
dynamical suppression in the case of approximate chiral 
symmetry as it is realised in Nature. The cross section should be suppressed 
by a factor $m_\pi^4/M^4$, where $M$ is a mass scale characterising 
the scattering process. In the calculation of the process 
$\gamma p \to \pi^0 N^*$ in \cite{Berger:1999ca} 
that effect had not been properly taken into account. 
A numerical estimate suggests that due to chiral symmetry 
the cross section found there is reduced by a factor of at least about 50 
\cite{Donnachie:2005bu}, changing the prediction (\ref{predict}) to 
less than about $6\, \mbox{nb}$. That reconciles the 
theoretical expectation with the experimental upper 
bound (\ref{exbound}) of \cite{Adloff:2002dw}. 

The considerations that we have outlined here can also be 
applied to pion production in other diffractive processes. 
An example that is relevant at the LHC and a future 
ILC is the quasidiffractive reaction $\gamma \gamma \to \pi^0 \pi^0$ 
at high energies. Also this reaction is at high energies mediated 
by odderon exchange. Here an even stronger suppression due to 
approximate chiral symmetry is expected.

\end{document}